\documentclass[prl,showpacs,twocolumn]{revtex4}
\usepackage{bm}
\usepackage{amsmath}
\usepackage{amssymb}
\usepackage{epsfig}

\newcommand{\eqb}{\begin{eqnarray}}
\newcommand{\eqe}{\end{eqnarray}}

\newcommand{\pd}{\partial}

\begin{document}
\title{Quantum switching at a mean-field instability of a Bose-Einstein condensate in an optical lattice}
\author{ V. S. Shchesnovich$^1$ and V. V. Konotop$^{2}$ }
\affiliation{ ${}^1$Centro de Ci\^encias Naturais e Humanas, Universidade
Federal do ABC, Santo Andr\'e,  SP, 09210-170 Brazil,\\
$^2$Centro de F\'isica Te\'orica e Computacional, Universidade de Lisboa, Complexo
Interdisciplinar, Avenida Professor Gama Pinto 2, Lisboa 1649-003, Portugal;
Departamento de F\'isica, Faculdade de Ci\^encias, Universidade de Lisboa, Campo
Grande, Ed. C8, Piso 6, Lisboa 1749-016, Portugal}

\begin{abstract}

It is shown that  bifurcations of the mean-field dynamics of a Bose-Einstein
condensate can be related with the quantum phase transitions of the original
many-body system. As an example we explore the intra-band tunneling in the
two-dimensional  optical lattice. Such a system allows for easy control by the
lattice depth as well as for macroscopic visualization of the phase transition. The
system manifests switching between two selftrapping states or from a selftrapping
state to  a superposition of the macroscopically  populated selftrapping states
with the step-like variation of the control parameter about the bifurcation point.
We have also observed the magnification of the microscopic difference between the
even and odd number of atoms   to a macroscopically distinguishable dynamics of the
system.

\end{abstract}
 \maketitle

\emph{Introduction.-} Since the very beginning of the quantum mechanics its
relation to the classical dynamics constitutes one of the central questions of the
theory. Dependence of the energy levels distribution  on the  type of dynamics of
the corresponding classical system~\cite{pioneer}, in general, and the quantum
system response to variation of the bifurcation parameters controlling the
qualitative changes of the classical behavior~\cite{responce} are among the major
issues~\cite{QuantChaos}. One of the main tools in studies of the quantum-classical
correspondence is the WKB approximation, where, loosely speaking, the Planck
constant $\hbar$ is regarded as a small parameter.

On the other hand, for a $N$-boson system  the limit $N\to\infty$ at a constant
density, leading to the mean-field approximation, can also be understood as a
semiclassical limit. This latter approach has  received a great deal of attention
during the last decade~\cite{PS}, due its high relevance to the theory of
Bose-Einstein condensates (BECs), many properties of whose dynamics are remarkably
well described  within the framework of the mean-field models~\cite{mean-field}.
More recently, it was shown \cite{SK1,SK2} that the mean-field description of a
few-mode $N$-boson system can be recast in a form similar to the WKB approximation
for a discrete Schr\"odinger equation \cite{Braun}, emergent for the coefficients
of the wavefunction expansion in the associated Fock space,  where $1/N$ plays the
role similar to that of the Planck constant in the conventional WKB approximation.

The mean-field equations of a system of interacting bosons are nonlinear, hence,
they naturally manifest many common features of the nonlinear dynamics, including
bifurcations of the stationary solutions caused by variation of the system
parameters. One of the well studied examples is a boson-Josephson
junction~\cite{Legg}, which can show either  equally populated (symmetric) or
strongly asymmetric states, characterized by population of only one of the sites
(the well known phenomenon of selftrapping~\cite{DK}). Now, exploring parallels
between the semiclassical approach and the mean-field approximation one can pose
the natural question: {\em what changes occur in a manybody system  when a control
parameter crosses  an instability (e.g. bifurcation) point of the limiting
mean-field system?}

In the present Letter we   give a partial answer  showing that one of
the possible scenarios is  the quantum phase transition of the second
type, associated with  the switching  of the wave-function in the Fock space
between the ``coherent" and ``Bogoliubov"  states  possessing distinct features.
Considering a flexible (time-dependent) control parameter, we have also found a
strong sensitivity of the system to the parity of the total number of atoms $N$,
showing  parity-dependent structure of the energy levels and the macroscopically
different dynamics for different parity of $N$. Observation of the discussed
phenomena is  feasible  in the experimental setting available nowadays.

\emph{Quantum and mean-field models.-} We consider the nonlinearity-induced
intra-band  tunneling of BEC between the two high-symmetry $X$-points of a
two-dimensional square optical lattice (OL). The process is described by the two
mode boson Hamiltonian (see \cite{SK1}  for the details)
\begin{equation}
\hat{H} = \frac{1}{2N^2}\left\{n_1^2+n_2^2 +\Lambda\left[4n_1n_2+(b_1^\dag b_2)^2 +
(b_2^\dag b_1)^2 \right] \right\}.
\label{EQ1}
\end{equation}
where $b_j$ and $b_j^\dag$ are the annihilation  and creation operators of the two
$X$-states, $\Lambda$ ($0\leq \Lambda\leq 1$) is the lattice parameter easily
controllable by variation of the lattice depth (or period). The Schr\"odinger
equation for the BEC in a state $|\Psi\rangle$ reads \mbox{$ih\pd_\tau |\Psi\rangle
= \hat{H}|\Psi\rangle$}, where  $h = 2/N$ and   $\tau = (2g\rho/\hbar)t$, with $g =
4\pi\hbar^2 a_s/m$  and the atomic density $\rho$. The link with the semiclassical
limit is evident for the Hamiltonian in the form (\ref{EQ1}):  the Schr\"odinger
equation written in the Fock basis,
$\displaystyle{|k,N-k\rangle=\frac{(b_1^\dagger)^k(b_2^\dagger)^{N-k}}{\sqrt{k!(N-k)!}}|0\rangle}$,
depends only on the relative populations  $k/N$ and \mbox{$(N-k)/N$}, while $h$
serves as an effective ``Planck constant''.

Hamiltonian (\ref{EQ1}) represents a nonlinear version of the well-known
boson-Josephson model (see, e.g. \cite{Legg,GO}), where unlike in the previously
studied models the states are coupled by the exchange of {\em pairs} of atoms. This
is a fairly common situation for systems with four-wave-mixing, provided by the
two-body interactions involving four bosons. The exchange of the  bosons by pairs
results in the coupling of the states with the same parity of the population and is
reflected in the double degeneracy of all $(N+1)/2$ energy levels for odd $N$,  due
to the symmetry relation $\langle 2k,N-2k|\Psi_1\rangle = \langle
N-2k,2k|\Psi_2\rangle$. For even $N$ the energy levels show quasi degeneracy (see
below).

The  mean-field limit of the system (\ref{EQ1}) can be formally obtained   by
replacing the  boson operators $b_j$ in (\ref{EQ1}) by the c-numbers $b_1\to
\sqrt{Nx}e^{i\phi/4}$ and $b_2\to \sqrt{N(1-x)}e^{-i\phi/4}$, what gives the
classical Hamiltonian~\cite{SK1}
\eqb
\mathcal{H} = x(1-x)\left[2\Lambda-1 + \Lambda\cos\phi\right] +\frac12,
\label{EQ9}
\eqe
where $x=\langle n_1\rangle/N$  is the  population density and $\phi=\arg\langle
(b_2^\dag)^2 b_1^2\rangle$ is the relative phase. $\mathcal{H}$ possesses  two
stationary points describing equally populated $X$-states: the classical energy
maximum $P_1=(x=\frac 12, \phi=0)$ and minimum $P_2=(x=\frac 12, \phi=\pi)$.
$P_1$ is dynamically stable in the domain $\Lambda>\Lambda_c= \frac13$. For
$\Lambda<\Lambda_c$ it looses its stability, and another set of stationary points
$x=1$ ($S_1$) and $x=0$ ($S_2$) appears, which is a fairly general situation in
nonlinear boson models. The appearing  solutions describe the symmetry breaking
leading to selftrapping.

\emph{Energy levels near the critical point.-} To describe the  spectrum of the
Hamiltonian (\ref{EQ1}) in the vicinity of the critical value $\Lambda_c$ we
rewrite $\hat{H}$  in terms of the operators $a_{1,2} = (b_1 \mp ib_2)/\sqrt{2}$
\begin{equation}
\hat{H}=\hat{H}_0+\left(1 -
\frac{\Lambda}{\Lambda_c}\right)\hat{V}+ {\cal E}(\Lambda) ,\,\,\, {\cal E}(\Lambda)=\frac{\Lambda+1}{4}+
\frac{\Lambda}{2N}
\end{equation}
where $ \hat{H}_0 =  \frac{2\Lambda}{N^2} a^\dag_1a_1a^\dag_2 a_2$ and $ \hat{V}=
\frac{1}{4N^2} \left(a_1^\dag a_2 +a_2^\dag a_1\right)^2$. At the critical point
the  energy spectrum is determined by  $\hat{H}_0$:  \mbox{$E_m =
\frac{2\Lambda_c}{N^2}m(N-m) +  {\cal E}(\Lambda_c)$}, where $m$ is the occupation
number corresponding to the operator $a_1^\dagger a_1$. The spectrum of $\hat{H}_0$
is doubly degenerate (except for the top level for even $N$) due to the symmetry
$m\to N-m$. The ground state energy is ${\cal E}_{min}(\Lambda_c)=E_0=E_{N}$, while
the top energy level has  $m = N/2$ for even $N$ and  $m = (N\pm1)/2$ for odd $N$.
Restricting ourselves to even number of bosons we get ${\cal
E}_{max}(\Lambda_c)=\frac12 +\frac{\Lambda_c}{2N}$.

\begin{figure}[htb]
\begin{center}
\epsfig{file=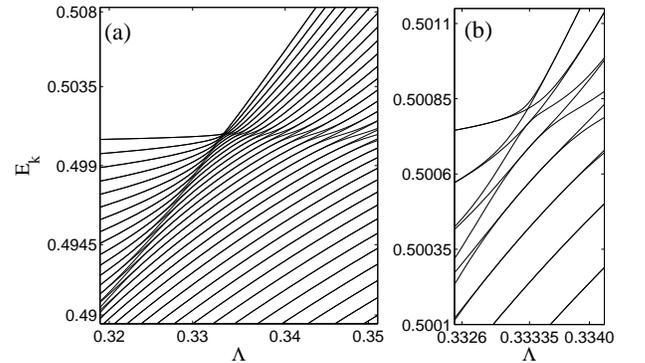,width=0.45\textwidth} \caption{(a) The energy levels
of $\hat{H}$  for $N=200$   and (b) a detailed picture in the vicinity of $\Lambda_c$.
The classical energy lines of the mean-field fixed points $P_1$ and $S_2$ are
visibly formed. The top energy  levels for  sufficiently large
$|\Lambda-\Lambda_c|$ are quasi-degenerate with the inter-level distances
indistinguishable on the scale of the figure (see the discussion in the text
below). }
\label{Fig1}
\end{center}
\end{figure}

Now consider small deviations of $\Lambda$  from the bifurcation point $\Lambda_c$.
To this end, for a fixed $N$, one can use the basis consisting of the degenerate
eigenstates of $\hat{H}_0$:
$\displaystyle{|E_m,j\rangle=\frac{(a_j^\dagger)^m(a_{3-j}^\dagger)^{N-m}}{\sqrt{m!(N-m)!}}|0\rangle}$,
$j=1,2$, $m=0,...,\frac N2$. The conditions for $\hat{V}$ to be treated as a
perturbation depend on $m$ as is seen  from the diagonal matrix
elements:
\begin{eqnarray}
\label{E_aver}
\langle E_m,j|\hat{V}|E_m,j\rangle = \frac{1}{4N}
+\frac{m}{2N}\left(1-\frac{m}{N}\right).
\end{eqnarray}
At the lower levels ($m\ll N/2$) the energy gaps between the degenerate subspaces
and the perturbation both scale as $\Delta E \sim N^{-1}$,  hence the condition of
applicability is $|\Lambda-\Lambda_c|\ll1$ and  the lower energy subspaces acquire
simple shifts. At the upper energy levels ($m\sim N/2$) the above energy gaps
behave as $\Delta E \sim N^{-2}$. Since $\langle \hat{V}\rangle \sim 1$ in this
case, the perturbation theory is applicable only in an interval of $\Lambda$ of the
size on the order of $N^{-2}$. There is a dramatic transition in the energy levels,
e.g. Fig.~\ref{Fig1}(b) shows the exchange  of the double degeneracy of the top
levels for even $N$ in this $N^{-2}$-small interval of $\Lambda$. By considering
the phase of
\eqb
\langle E_m,j|(b_2{\!}^\dagger b_1)^2|E_m,j\rangle = -\frac{N^2}{4}+\frac N4+\frac
32m\left(  N-m\right),
 \label{phase}\eqe
it is  easy to verify  that the upper and lower eigenstates correspond,
respectively, to the mean-field stationary points $P_1$ ($\phi=0$) and $P_2$
($\phi=\pi$).

\emph{Spectrum in the limit $N\to\infty$. Coherent states and selftrapping states.
-} For $ \Lambda_c^{-1}-\Lambda^{-1} \gg N^{-2} $ the quantum states corresponding
to $P_1$ can be obtained by  quantizing the local classical Hamiltonian
(\ref{EQ9}), i.e. by expanding it with respect to $ x-1/2$  and  $\phi$ and setting
$\phi = -ih\frac{\partial}{\partial x}$ (see also Ref. \cite{ST}; on this way one
looses the term of order $1/N$ in $\mathcal{E}_{max}$). The ``wave function''
$\psi(x)=\sqrt{N}C_k \equiv \sqrt{N}\langle k,N-k|\psi\rangle$ satisfies
\eqb
\left[\frac{\Lambda h^2}{8}\frac{\partial^2}{\partial  x^2}  +(3\Lambda-1)
\left(\frac 14- \left(x -\frac 12\right)^2\right)\right]\psi = E\psi .
\label{EQ11}
\eqe
Eq. (\ref{EQ11}) is the negative mass quantum oscillator problem with the frequency
$\omega^2=8\left(3-\frac{1}{\Lambda}\right)$. The respective descending energy
levels read $ E^{(\mathrm{top})}_n =  {\cal E}_{max} +
\frac14\left(\frac{\Lambda}{\Lambda_c}-1\right)  -
\frac{h\Lambda\omega}{4}\left(n+\frac12\right).$ The eigenfunctions  are localized
in the Fock space, e.g. the $n=0$ eigenfunction  is $ \psi_0(x) =
C\exp\left[-\frac{\omega}{2h}\left(x-\frac 12\right)^2\right]$.  In the original
discrete variable $x = k/N$, there are even and odd eigenstates $C_{2k}$ and
$C_{2k-1}$ related by the approximate symmetry $C_{n} \approx C_{n+1}$, hence the
energy levels are quasi doubly degenerate [c.f.  Fig. 1(b)].

The local  approximation  becomes invalid as \mbox{$\Lambda_c^{-1}-\Lambda^{-1}\sim
N^{-2}$} (the wave-function delocalizes).  The other set of the stationary points,
$S_{1,2}$, becomes stable for $\Lambda < \Lambda_c$ in the mean-field limit. In
this case, however, the phase $\phi$ is undefined. Let us first consider the full
quantum case, for example, the limit $\langle n_1 \rangle \ll N$ (i.e. the point
$S_2$). The resulting reduced Hamiltonian can be either easily derived in the Fock
basis or obtained by formally setting $b_2= N$ and retaining the lowest-order terms
in $b_1$ and $b_1^\dagger$:
\eqb
\hat{H}\approx \hat{H}_{S_2} = \frac12 +\frac{(2\Lambda-1)}{N} b_1{\!}^\dagger b_1
+ \frac{\Lambda}{2N}[(b_1^\dagger)^2 +b_1^2].
\label{EQ16}
\eqe
Hamiltonian (\ref{EQ16}) can be diagonalized by the Bogoliubov transformation
$c=\cosh (\theta) b_1 - \sinh(\theta) b_1{\!}^\dagger$, where
$\theta=\theta(\Lambda)>0$ is determined from $\tanh(2\theta) = \Lambda/(1-
2\Lambda)$. We get
\eqb
\hat{H}_{S_2} = -\frac{\Lambda}{N\sinh(2\theta)} c^\dagger c +\frac{\Lambda
\tanh\theta}{2N}+ \frac12.
\label{EQ18}
\eqe
Thus $c^\dag c$ gives the number of  negative-energy quasi-particles over  the
Bogoliubov (squeezed) vacuum solving \mbox{$c|\mbox{vac}\rangle = 0$}. In the
atom-number basis $|\mbox{vac}\rangle$ is a superposition of the Fock states with
$C^{(vac)}_{2k} = \tanh^k(\theta)\sqrt{(2k)!}/(2^k k!)C_0$ and $ C^{(vac)}_{2k-1} =
0,$ ($C_0$ is a normalization constant).

\begin{figure}[htb]
\begin{center}
\epsfig{file=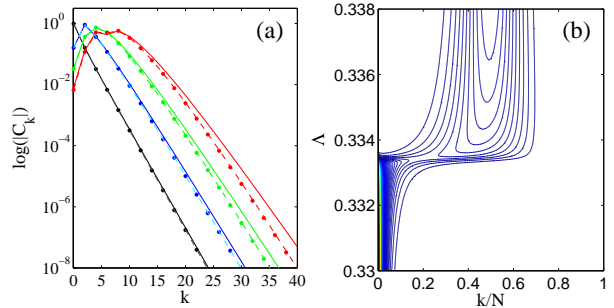,width=0.45\textwidth} \caption{(Color online)  (a)
Convergence of the four upper eigenstates of the Hamiltonian (\ref{EQ1}) to the
eigenstates of $\hat{H}_{S_2}$ (shown by dots) for for $N=100$ (solid lines) and $N
= 1000$ (dashed lines), for $\Lambda=\Lambda_c-0.1$. (b) The contour plot of the
state corresponding to the top energy level in the vicinity of $\Lambda_c$ for $N =
200$. }
\label{Fig2}
\end{center}
\end{figure}

The validity condition of the approximation (\ref{EQ16}), given by $ \langle
\hat{n}\rangle, \Delta n \ll N$, can be rewritten in the form
$\tanh^{-2}(2\theta)\gg 1+N^{-2} $, what is the same as
$\Lambda^{-1}-\Lambda_c^{-1}\gg N^{-2}$.     In this case, the eigenstates of
(\ref{EQ16}) are well-localized in the atom-number Fock space, i.e. the
coefficients $C_{2k}$ decay fast enough. The  condition for this excludes the same
small interval as in the perturbation theory, hence the transition between the
coherent states and the selftrapping (Bogoliubov) states occurs on the interval of
$\Lambda$ of order of $N^{-2}$. The convergence of the eigenstates of
$\hat{H}_{S_2}$ to that of the full Hamiltonian (\ref{EQ1}) turns out to be
remarkably fast as it is shown in Fig.~\ref{Fig2}(a).  In Fig.~\ref{Fig2}b the
dramatic deformation of the top energy eigenstate of $\hat{H}$ (corresponding to
the $S_2$-$P_1$ transition) about the critical $\Lambda_c$ is shown. Finally, we
note that for even $N$ the quasi double degeneracy of the energy levels for
$\Lambda^{-1}-\Lambda_c^{-1}\gg N^{-2}$ (c.f. Fig. 1(b))  is due to the exchange
symmetry between $S_1$ and $S_2$ resulting in equal energy levels of the
Hamiltonians $\hat{H}_{S_1}$ and $\hat{H}_{S_2}$.

\medskip

In the mean-field description of the  stationary point $S_2$  the associated
Hamiltonian is defined by replacing  the boson operators in Eq. (\ref{EQ16}) by the
c-numbers $b_1 = \sqrt{N}\alpha$ and $b_2 = \sqrt{N}\beta$.  Using
$|\alpha|^2+|\beta|^2=1$ and fixing the irrelevant common phase by setting $\beta$
real we get the dynamical variables $\alpha$ and $\alpha^*$ and the classical
Hamiltonian in the form $\mathcal{H}_{S_2} = \frac12
+\frac12(1-|\alpha|^2)\{2(2\Lambda-1)|\alpha|^2+\Lambda[\alpha^2 +(\alpha^*)^2]\}$,
from which the stability of the point $S_2$ ($\alpha=0$) for $\Lambda<\Lambda_c$
follows.

Thus, {\em  the passage  through the bifurcation point $\Lambda_c$ of the
mean-field model, corresponds to the phase transition in the quantum many-body
system on an interval of the control parameter scaling as $N^{-2}$ and reflected in
the deformation of the spectrum and dramatic change of the system wave-function in
the Fock space. The described change of the system is related to the change of the
symmetry of the atomic distribution, and thus it is the second order phase
transition}.

In our case this scenario  corresponds to loss of stability of the selftrapping
solutions $S_1$ and $S_2$ and appearance of the stable stationary point $P_1$. In
the quantum description this happens by  a set of avoided crossings  of the top
energy levels (and splitting of the quasi-degenerate energy levels  for even $N$)
as the parameter $\Lambda$ sweeps the small interval on the order of $N^{-2}$ about
the critical value $\Lambda_c $ (see Fig.~\ref{Fig1}).  For lower energy levels the
avoided crossings appear along the two straight lines approximating the classical
energies of the two involved stationary points: \mbox{$\mathcal{H}(P_1) = \frac 12
+\frac{\left(3\Lambda-1\right)}{4} $} (for $\Lambda<\Lambda_c$) and ${\cal E}_{max}
= \frac 12$ ($\Lambda>\Lambda_c$), see Fig. \ref{Fig1}.

\emph{Dynamics of the phase transition.-} Let us see how the quantum phase
transition shows up in the system dynamics when $\Lambda$ is time-dependent. The
selftrapping states $S_1$ and $S_2$, eigenstates of the Hamiltonian  (\ref{EQ1}),
correspond to occupation of just one of the  $X$-points. Such an initial condition
can be experimentally created by switching on a moving lattice with
$\Lambda<\Lambda_c$ (see e.g.~\cite{SK2}). As the lattice parameter $\Lambda(\tau)$
passes the critical value from below, the selftrapping states are replaced by the
coherent  states with comparable average occupations of the two $X$-points.

\begin{figure}[htb]
\begin{center}
\epsfig{file=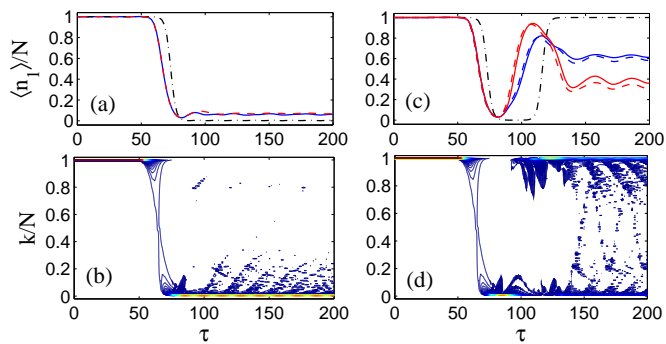,width=0.495\textwidth} \caption{(Color online)  The average
population densities $\langle n_1 \rangle/N$, (a) and (c), and the atom-number
probabilities $|C_k|^2 $, (b) and  (d), for  $\Lambda(\tau)=\Lambda_1 +
(\Lambda_2-\Lambda_1)\left[\tanh\left(\tau-\tau_1 \right) - \tanh\left( \tau-\tau_2
\right)\right]/2$. The corresponding classical dynamics  is shown by the dash-dot
lines in (a) and (c). Here $\Lambda_1=0.25<\Lambda_{cr}$, $\Lambda_2 =
0.5>\Lambda_{cr}$, $\tau_1 = 50$ and  $\tau_2 = 85$ (a) with  $N = 500$ and $501$
(indistinguishable), while in (c) $\tau_2 = 135$  with $N = 500$ and $400$ (the
upper solid and dashed lines) and $N = 501$ and $401$ (the lower lines).  The
initial state is $|\mathrm{vac}\rangle$ of  $H_{S_1}$, but using $|N,0\rangle$
gives a similar picture.}
\label{Fig3}
\end{center}
\end{figure}

A more intriguing  dynamics  is observed when $\Lambda(\tau)$ is a smooth step-like
function between  $\Lambda_1$ and $\Lambda_2$ such that $\Lambda_1 < \Lambda_c<
\Lambda_2$. In this case, the system dynamics and the emerging states dramatically
depend also on parity of the number of atoms. For  fixed $\Lambda_{1,2}$  the
system behavior crucially depends on the time that $\Lambda(\tau)$ spends  above
$\Lambda_c$. More specifically, one can identify two distinct scenarios, which can
be described as a switching dynamics between the selftrapping states  at the two
$X$-points, Fig.~\ref{Fig3}(a),(b) or dynamic creation of the superposition of
macroscopically distinct states, well approximated by
\mbox{$\sum\limits_{k<k_m}\left(C_k|k,N-k\rangle + C_{N-k}|N-k,k\rangle\right)$}
with a small $k_m/N$, Fig.~\ref{Fig3}(c),(d) (where $k_m/N \approx 0.2$).  In the
case of macroscopic superposition the dynamics  shows anomalous dependence on
parity of $N$, i.e. showing the same behavior for large $N$ of the same parity but
macroscopically distinct behavior for $N$ and $N+1$,  Fig.~\ref{Fig3}(c). Note that
the mean-field dynamics is close to the quantum one in the switching case,
Fig.~\ref{Fig3}(a), while it is dramatically different in the superposition case,
Fig.~\ref{Fig3}(c).

\medskip

To estimate the physical  time scale, $t \equiv t_\mathrm{ph}\tau =
\frac{m d^2 \ell_\perp}{8\pi\hbar a_s \mathcal{N}_\mathrm{pc}}\tau$,  we assume
that a condensate of ${}^{87}$Rb atoms is loaded in a square lattice with the mean
density of $\mathcal{N}_\mathrm{pc} = 20$ atoms per cite.  If the lattice constant
$d = 2\,\mu$m and the oscillator length of the tight transverse  trap (to assure
the two-dimensional approximation) $\ell_\perp = 0.1\,\mu$m, then $t_\mathrm{ph}
\sim 0.2$ ms and the time necessary for the creation of the macroscopic
superposition of Fig. \ref{Fig3}(c),(d) is about $20$ ms.

\emph{Conclusion.-} We have shown that behind the mean-field instability in the
intra-band tunneling of BEC in an optical lattice is a quantum phase transition
between macroscopically distinct states, giving a macroscopic magnification of the
microscopic quantum features of the system. A spectacular demonstration of this is
the dynamic formation of  the superposition of  macroscopically distinct states,
which, besides being responsible for the difference between  the mean-field and
quantum dynamics (see also recent Ref. \cite{WT}), shows also an anomalous
dependence on parity of BEC atoms reflecting distinct energy level structure for
even and odd number of atoms.

\acknowledgments   The work of VSS was supported by  the FAPESP of Brazil.


\begin{thebibliography}{99}

\bibitem{pioneer} I. C. Percival. J. Phys. B {\bf 6}, L229 (1973);
M. V. Berry and M. Tabor, Proc. R. Soc. Lond. A {\bf 356}, 375 (1977).

\bibitem{responce}
P.  Pechukas, Phys. Rev. Lett. {\bf 51}, 943 (1983);
T. Yukawa, Phys. Rev. Lett. {\bf 54}, 1883 (1985).

\bibitem{QuantChaos} see e.g. K. Nakamura, {\em Quantum Chaos}
(Cambridge University Press, 1993); F. Haake, {\em Quantum Signatures of Chaos} (Springer-Verlag Berlin Heidelberg, 2001)

\bibitem{PS} L. Pitaevskii and S. Stringari, {\em Bose-Einstein Condensation} (Clarendon Press, Oxford, 2003)

\bibitem{mean-field} C. W. Gardiner, Phys. Rev. A {\bf 56}, 1414 (1997);
 Y. Castin and R. Dum, Phys. Rev. A {\bf 57}, 3008 (1998).

\bibitem{SK1} V. S. Shchesnovich and V. V. Konotop, Phys. Rev. A \textbf{75}, 063628
(2007).

\bibitem{SK2} V. S. Shchesnovich and V. V. Konotop, Phys. Rev. A \textbf{77}, 013614
(2008).

\bibitem{Braun} P. A. Braun, Rev. Mod. Phys. \textbf{65}, 115 (1993).

\bibitem{Legg} A. J. Leggett, Rev. Mod. Phys. \textbf{73,} 307 (2001).

\bibitem{DK} V. M. Kenkre and D. K. Campbell, Phys. Rev. B {\bf 34},  4959 (1986).


\bibitem{GO} S. Raghavan, A. Smerzi, S. Fantoni, and S. R. Shenoy, Phys. Rev. A {\bf 59}, 620 (1999);
R. Gati and M. K. Oberthaler, J. Phys. B: At. Mol. Opt. Phys. \textbf{40,} R61.
(2007).

\bibitem{ST} V. S. Shchesnovich and M. Trippenbach, arXiv: cond-mat/08040234; Phys. Rev. A \textbf{78}, 023611 (2008).

\bibitem{WT} C. Weiss and N. Teichmann, Phys. Rev. Lett. \textbf{100,} 140408 (2008).

\end{thebibliography}
\end{document}